\documentclass[journal]{IEEEtran}
\usepackage{graphicx,epsfig}
\usepackage{epstopdf}
\usepackage{amsmath}
\usepackage{cases}
\newtheorem{thm}{Theorem}%[section]

\ifCLASSINFOpdf
\else
\fi
\begin{document}
\title{Energy Efficient Optimization of Wireless-powered 5G Full Duplex Cellular Networks: A Mean Field Game Approach}

\author{Xiaohu~Ge,~\IEEEmembership{Senior Member,~IEEE,}
        Haoming~Jia,
        ~Yi~Zhong,~\IEEEmembership{Member,~IEEE,}
        Yong~Xiao,~\IEEEmembership{Senior Member,~IEEE,}
        Yonghui~Li,~\IEEEmembership{Senior Member,~IEEE,}
        Branka~Vucetic~\IEEEmembership{Fellow,~IEEE,}% <-this % stops a space

}

% note the % following the last \IEEEmembership and also \thanks - branka.vucetic@sydney.edu.au


\maketitle

\begin{abstract}
This paper studies the power allocation of an ultra-dense cellular network consisting of multiple full-duplex (FD) base stations (BSs) serving a large number of half-duplex (HD) user equipments (UEs) located in a wide geographical area. Each BS consists of a baseband unit (BBU) that is responsible for signal processing and base station control, and a radio remote unit (RRU) that corresponds to a radio transceiver remotely built closer to the UEs. We consider a wireless-powered cellular network in which the BBU can periodically charge the RRU. We model the energy efficiency and coverage optimization problem for this network as a mean field game. %Optimal transmit powers of the BSs have been derived. %
We consider the weighted energy efficiency of the BS as the main performance metrics and evaluate the optimal strategy that can be adopted by the FD BSs in an ultra-dense network setting.
Based on the interference and network energy efficiency models of the mean field game theory,
Nash equilibrium of our proposed game is derived. % for 5G cellular networks with full duplex transmissions.
Utilizing solutions of Hamilton-Jacobi-Bellman (HJB) and Fokker-Planck-Kolmogorov (FPK) equations, a new power transmission strategy is developed for the BSs to optimize the energy efficiency of 5G cellular networks with full duplex transmissions. Simulation results indicate that the proposed strategy not only improves the energy efficiency but also ensures  the average network coverage probability converges to  a stable level. % for 5G cellular networks with full duplex transmissions.
\end{abstract}


% \fi
\IEEEpeerreviewmaketitle

\section{Introduction}

\IEEEPARstart{F}{u}{l}{l}-duplex has been considered as one of the key technologies to improve the spectrum utilization of 5G networks \cite{Wang4,zhang111,zhang112,liqiang,zhang115}. In the areas with dense traffic, such as the shopping malls and the stadiums, the spectrum resource is insufficient for the high-throughput applications \cite{YZhong18ComMag}. Thus, full-duplex is necessary to further improve the spectrum utilization and network capacity.
However, %since the traffic can be highly unbalanced for downlink and uplink transmissions in 5G cellular networks and due to the advantage of simplicity and more controllable interference, the
half-duplex technology is still widely used for user equipments (UEs) in most existing applications due to its simplicity and reduced interference.

When the 5G networks are densely deployed in an urban environment, the base stations (BSs) will have to be installed at inaccessible locations such as at the top of the building and lamp post. One possible solution is to divide the BS into two physically separated entities: the baseband unit (BBU) installed at enclosed space such as street cabinet that is responsible for signal processing and base station control, and the radio remote unit (RRU) with a radio transceiver that can be remotely built at the top of the hill lamp post. One of the main challenges for this solution is to support sufficient and reliable power supply to the RRU. In particular, the RRU can be installed at some locations inaccessible to the power grid.

In this paper, we assume that the BBU can wirelessly send the power to the RRU with the wireless power transfer technology; thus, the battery of RRU can be periodically charged, see \cite{Ozel5}\cite{Cheng3}.
%The energy harvesting technology is proposed to periodically charge the battery of RRU \cite{Ozel5}\cite{Cheng3}.
In particular, each RRU can have a battery that can be periodically charged by the BBU. The transmission power of RRU is constrained by the battery volume. Low battery level may increase the outage probability. Therefore, it is a great challenge for optimizing the transmission power strategy of RRU to not only keep the coverage of the network at a satisfying level but also improve the energy efficiency of 5G cellular networks \cite{Zhong17TWC} considering both downlinks and uplinks with full duplex transmissions \cite{Ge11}.
To improve the spectrum and energy efficiency, the full-duplex transmission and energy harvesting technologies are emerging as promising technologies for 5G wireless communication systems \cite{Duarte6}, which we will introduce respectively in the following discussions.

The full-duplex transmission technology can be used to transmit and receive wireless signals in the same frequency at same time , thus improving the spectrum efficiency of wireless communications \cite{Huang7}. However, the self-interference is an inevitable problem \cite{Zhang8}, which has been explored extensively in the literature. For example, several self-interference cancellation technologies were introduced in \cite{Zhang9,Sim1,Liu1}. In particular, power allocation problem has been formulated as a non-convex optimization problem \cite{Li10}. Solutions have been presented to to improve the spectrum and energy efficiency for full-duplex communication systems with massive MIMO. Algorithms based on path-following method have been developed for jointly optimizing the energy harvesting time allocation and beamformers to maximize the sum rate and energy efficiency of a full-duplex transmission system \cite{Nguyen12}. The authors in \cite{Wen13} proved that energy consumption is a quasi-concave function of spectrum utilization in a cellular network with full-duplex transmissions. An optimal energy efficient oriented resource allocation algorithm have been developed under spectrum efficiency constraints \cite{Wen13}.

The energy harvesting technologies have been proposed to collect energy from the ambient environment for extending the battery life of wireless networks \cite{Mao14}. When small cell BSs harvest energy from external power sources, a discrete single-controller discounted two-player stochastic game was formulated to investigate the problem of downlink power control in a two-tier microcell-small cell network \cite{Thuc15}. By modelling the energy harvesting and consumption as a probability distribution,  a bandit framework was proposed to solve the user association problem in a distributed manner \cite{Maghsudi16}. To overcome the uncertainty of the environmental conditions that affects the harvesting energy,  two online energy trading approaches,  one decentralized and one centralized were proposed to minimize the nonrenewable energy consumption in a multi-tier cellular network \cite{Reyhanian17}.

Based on the energy harvesting and full-duplex transmission technologies,  the cell association and BS power allocation scheme were proposed to optimize the energy efficiency of cellular networks \cite{Chen18}. By modelling the battery dynamics of an energy harvesting small cell BS as a discrete-time Markov chain and considering a practical power consumption model,  a tractable model was proposed to optimize the energy efficiency of heterogeneous cellular networks \cite{Yu19}. With the massive MIMO and millimeter wave transmission technologies adopted for 5G wireless communications, small cell BSs have to be densely deployed in 5G cellular networks \cite{Ge1}.

%Since 5G cellular networks will comprise of a large number of small cell BSs \cite{Ge11,Pitaval11,zhang113,Lin},  the mean field game theory is a natural tool to investigate the problem of 5G cellular networks.

The mean field game is a mathematic framework focusing on the strategic decision making in a very large population of small interacting agents \cite{Lasry20}\cite{Huang21}\cite{Huang22}. Moreover,  the mean field game theory has been used for studying wireless networks \cite{Tembine23}\cite{Meri25}\cite{Meri27}. The interaction between the primary user and a large number of secondary users was formulated as a hierarchical mean field game \cite{Tembine23}. Considering the energy availability and the impact of channel fading, an energy constraint power control problem was formulated as a mean field game \cite{Meri25}. Furthermore, the mean field game is utilized to model the energy efficient wireless networks \cite{Sama28}\cite{Yang29}. By bringing notions of the Lyapunov optimization and mean-field theory, a new approach has been proposed to jointly improve the power control and the user scheduling in an ultra-dense small cell network \cite{Sama28}. Based on a mean-field game theoretic approach, a novel energy and interference aware power control policy was proposed to optimize the energy efficiency of ultra-dense device-to-device (D2D) Networks \cite{Yang29}.
However, in all the aforementioned studies, only simple scenarios, such as BSs with single antenna or half duplex technologies, were analyzed by the mean field game theory, and a simple power control scheme was developed for traditional cellular networks.

The network energy efficiency adopting full duplex and wireless power transfer technologies is a promising solution to meet the requirements of the future wireless networks, which is still relatively under explored. % surprisingly rare in the existing literature.
Motivated by the above gaps, in this paper we derive the network energy efficiency of 5G cellular networks adopting full duplex and wireless power transfer technologies. The contribution of this paper is summarized as follows.

\begin{enumerate}
\item Considering the unbalance between downlinks and uplinks and periodically battery charging at RRUs, the energy efficiency of BSs with full duplex transmissions is formulated to evaluate the performance of RRU transmission power.
\item Based on the mean field game theory, the interference and network energy efficiency of 5G cellular networks with full duplex transmissions are derived for performance analysis.
\item Based on the solution of mean field game equations, a new RRU transmission power strategy, {\em i.e.,} the equilibrium of mean field game (EMFG) algorithm is developed to optimize the energy efficiency of 5G cellular networks with full duplex transmissions.
\item Simulation results show that the network energy efficiency with the EMFG algorithm is larger than the network energy efficiency with the fixed RRU transmission power algorithm in 5G cellular networks. Moreover, the cellular network with the EMFG algorithm is helpful for keeping the average network coverage probability at a stable level.
\end{enumerate}

The remainder of this paper is outlined as follows. Section II describes the system model. Considering the unbalance between downlinks and uplinks and periodically battery charging at RRUs, the  energy efficiency of the single BS with full duplex transmission is formulated in Section III. Based on the mean field game theory, the interference and network energy efficiency are derived for cellular networks with full duplex transmission in Section IV. In Section V, a new RRU transmission power algorithm is developed to optimize the energy efficiency of 5G cellular networks. Furthermore, simulation results are analyzed and discussed in Section VI. Finally, Section VII concludes this paper.

\section{System Model}
It has been verified by many existing works \cite{Andrews39, Zhong1, Zhong2, Zhong3} that the locations of the UEs as well as SBSs follow the stochastic geometry models. We follow the same model in this paper and assume the locations of the UEs and BSs can be modeled as two independent Poisson point processes with intensities
$\lambda_{\rm UE}$ and $\lambda_{\rm BS}$, respectively.
We consider a cellular network consisting of a number of BSs as illustrated in Fig. 1(a).
Each BS is composed of a BBU and an RRU. The BBU is used for baseband processing and
responsible for the communication through the physical interface such as processing and forwarding voice and data requests to the UE. RRU is an RF processing unit that can be installed separately to extend the service coverage of the BS. The BBU and RRU are connected by optical fiber links.
We assume the BBU is always supported by reliable power sources, e.g., power grid. The RRU, however, is located in an inaccessible locations (e.g., lamp post, street cabinets, etc.) and can only supported by the power wirelessly transferred from the BS. Each RRU is installed with an energy storage device (battery or super-capacitor) with limited capacity ${\bar p}_T$.
Letting $p_{T}$ be the transmit power of the RRU associated with the $k$th BS ($k\in N^+$), we get $0 \le p_T \le {\bar p}_T$.
Since the transmit power of UEs is much less than that of the BSs, we ignore the interference from all UEs. We list the main notations adopted in this paper in Table I.

In this case,  the BBU is usually supplied by the traditional power lines. However,  the power supply of RRU is an issue considering the deployment problem of electricity lines in outdoor scenarios. Based on the solution in \cite{Ozel5}, in this paper every separated RRU is assumed to be equipped with a battery which can be charged by the BBU via the wireless power transfer technology. The basic structure of BBU and RRU is shown in Fig. 1(b). Considering the volume limit of battery,  the maximum transmission power of a RRU is configured as $\rm p_{{T-RRU}_{max}}$. The transmission power of the RRU at the $k$th  BS is denoted by $\rm p_{{T-RRU}_\emph{k}}$, $k=1, 2, 3,...$. The value range of $\rm p_{{T-RRU}_\emph{k}}$ is given by $0\leq \rm p_{{T-RRU}_\emph{k}}\leq \rm p_{{T-RRU}_{max}}$.
\begin{figure}[!t]%1
\centering
\includegraphics[width=0.5\textwidth]{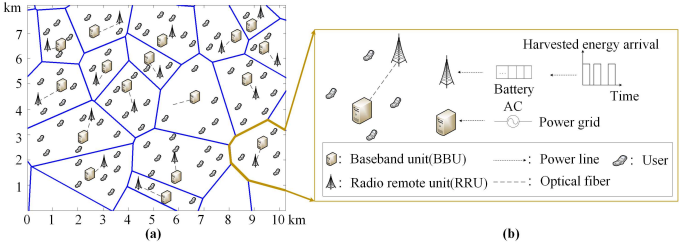}
\caption{Cellular networks with wireless power transfer}
\label{Fig. 1.}
\end{figure}

We consider the BSs equipped with full-duplex RRU can serve different UEs in both uplinks (from BS to UE) and downlinks (from UE to BS) at the same time over the same frequency. The UEs can however only operate in half-duplex mode as illustrated in Fig. 2. In this paper we focus on optimizing the energy efficiency of cellular networks with full duplex and wireless power technologies in the time slot $t$, $0\leq t\leq T$, where $T$ is the battery charging period of RRUs.

\begin{figure}[!t]%2
\centering
\includegraphics[width=0.5\textwidth]{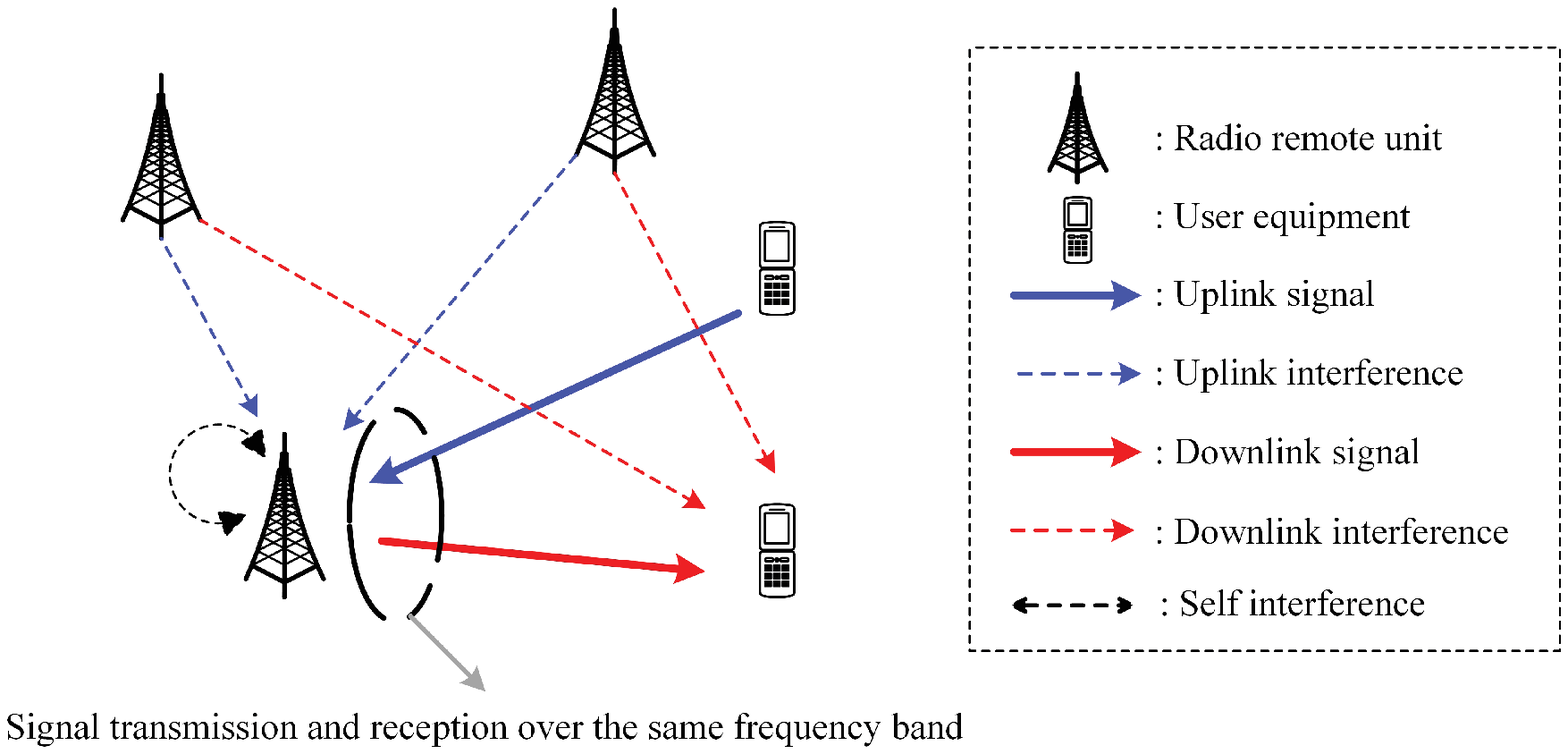}
\caption{Full duplex transmission scheme}
\label{Fig. 2.}
\end{figure}

\begin{table*}[!t]
\renewcommand{\arraystretch}{1.3}
\label{Table I}
\caption{\ Symbols and variables table}
\centering
\begin{tabular}{c|c}
\hline
    \textbf{Symbol and variable} & \textbf{Explanation}\\
    \hline
    $\rm p_{{T-RRU}_\emph{k}}$ & Transmission power of the RRU at the $k$th  BS \\
    $\rm SINR_{k_i}^{\rm DL}$ & SINR of downlink between $\rm {BS}_\emph{k}$ and $\rm UE_\emph{i}$\\
    $\rm r_{\emph{k}_\emph{i}}^{RRU-UE}$ & Distance between the RRU of $\rm {BS}_\emph{k}$ and $\rm {UE}_\emph{i}$ \\
    $\rm {SINR}_{\emph{k}_\emph{i}}^{UL}$ & SINR of uplink between $\rm {UE}_\emph{i}$ and $\rm {BS}_\emph{k}$\\
    $\rm r_{\emph{m}_\emph{k}}^{RRU-RRU}$ &  Distance between the RRU of interfering $\rm {BS}_\emph{m}$ and $\rm {BS}_\emph{k}$\\
    $\rm I_{\emph{k}}^{UL}$ & Aggregated interference from adjacent BSs except for $\rm {BS}_\emph{k}$\\
    $\rm U_{B{S_\emph{k}}}$ & Weighted energy efficiency of a cell associated with the $\rm {BS}_\emph{k}$\\
    $\rm EE_{\emph{k}_{\emph{i}_{DL}}}^{DL}$ & Energy efficiency of downlinks between $\rm {BS}_\emph{k}$ and $\rm {UE}_{\emph{i}_{DL}}$\\
    $\rm EE_{\emph{k}_{\emph{i}_{UL}}}^{UL}$ & Energy efficiency of uplinks between $\rm {BS}_\emph{k}$ and $\rm {UE}_{\emph{i}_{UL}}$\\
    $\rm e_\emph{k}$ & Residual volume of the battery used for a RRU at $\rm {BS}_\emph{k}$\\
    $\rm V_\emph{k}$ & Value function of the $\rm BS_\emph{k}$ \\
    $\rm I_{mean-RR{U_\emph{k}}}$ & Interference aggregated at the BS $\rm BS_\emph{k}$\\
    $\rm I_{mean-U{E_{\emph{i}_{DL}}}}$ & Interference aggregated at the user $\rm UE_{\emph{i}_{DL}}$\\
    $\rm EE_{\emph{k}}^{DL-mean}$ & Energy efficiency of downlinks in the mean field\\
    $\rm EE_{\emph{k}}^{UL-mean}$ & Energy efficiency of uplinks in the mean field\\
    $\rm m(\emph{t}, \emph{e})$ & Mean field of network energy efficiency\\
    $\rm U_{BS_\emph{k}}^{mean}(p_{T-RR{U_\emph{k}}})$ & Weighted energy efficiency of the $\rm BS_\emph{k}$ in the mean field\\
    \hline
\end{tabular}
\end{table*}

\section{Formulation of Energy Efficiency}
\subsection{SINR for Downlinks}
Consider a cellular network with the topology shown in Fig. 1(a),  the interference from the BSs and the wireless channel noise are included into the received signals at UEs. Therefore, the signal-to-interference-and-noise ratio (SINR) received at the $i$th UE in the time slot $t$ is given by,
\begin{equation}
\label{1}
{\rm SINR}_{k_i}^{\rm DL}(t) = \frac{\rm p_{T-RR{U_\emph{k}}}(\emph{t}) h_{\emph{k}_\emph{i}}^{RRU-UE}(\emph{t}) r_{\emph{k}_\emph{i}}^{RRU-UE}{(\emph{t})^{-\alpha}}}
{\rm I_{\emph{k}}^{DL}(\emph{t})+ N_0},
\end{equation}
where $\rm p_{T-RR{U_\emph{k}}}(\emph{t})$ is the transmission power of the $k$th BS, $\rm h_{\emph{k}_\emph{i}}^{RRU-UE}(\emph{t})$ is the small scale fading between the RRU of the $k$th BS and the $i$th UE, $\rm r_{\emph{k}_\emph{i}}^{RRU-UE}(\emph{t})$ is the distance between the RRU of the $k$th BS and the $i$th UE, $\rm \alpha$ is the path loss fading,  $\rm N_0$ is the additive noise received by UE $i$.
$\rm I_{\emph{k}}^{DL}(\emph{t})=\sum\limits_{\emph{m}\in N^{+},\emph{m}\neq \emph{k}}\rm p_{T-RR{U_\emph{m}}}(\emph{t})\cdot h_{\emph{m}_\emph{i}}^{RRU-UE}(\emph{t})\cdot
r_{\emph{m}_\emph{i}}^{RRU-UE}(\emph{t})^{-\alpha}$ is the interference from RRUs of the interfering BSs, $\rm p_{T-RR{U_\emph{m}}}(\emph{t})$ is the RRU transmission power of the interfering BS $\rm {BS}_\emph{m}$, $\rm h_{\emph{m}_\emph{i}}^{RRU-UE}(\emph{t})$ is the small scale fading between the RRU of interfering BS $\rm {BS}_\emph{m}$ and the UE $\rm {UE}_\emph{i}$, $\rm r_{\emph{m}_\emph{i}}^{RRU-UE}(\emph{t})$ is the distance between the RRU of the $m$th interfering BS and the $i$th UE.

\subsection{SINR of Uplinks}
Similarly, the SINR of uplink between the desired UE $\rm {UE}_\emph{i}$ and the BS $\rm {BS}_\emph{k}$ at the time slot $t$ is given by
\begin{equation}
\label{2}
\rm {SINR}_{\emph{k}_\emph{i}}^{UL}(\emph{t}) = \frac{p_{T-UE}(\emph{t})h_{\emph{k}_\emph{i}}^{RRU-UE}(\emph{t}) r_{\emph{k}_\emph{i}}^{RRU-UE}{(\emph{t})^{-\alpha}}}
{p_{T-RR{U_\emph{k}}}(\emph{t})\cdot h_{self}(\emph{t})+I_{\emph{k}}^{UL}(\emph{t})
+ N_0},
\end{equation}
where $\rm h_{self}(\emph{t})$ is the gain of the self-interference channels at RRUs, $\rm p_{T-RR{U_\emph{k}}}(\emph{t})\cdot h_{self}(\emph{t})$ is the self-interference at the BS $\rm {BS}_\emph{k}$, $\rm h_{\emph{m}_\emph{k}}^{RRU-RRU}(\emph{t})$ is the small scale fading between the RRU of interfering BS $\rm {BS}_\emph{m}$ and the RRU of desired BS $\rm {BS}_\emph{k}$, $\rm r_{\emph{m}_\emph{k}}^{RRU-RRU}(t)$ is the distance between the RRU of interfering BS $\rm {BS}_\emph{m}$ and the RRU of desired BS $\rm {BS}_\emph{k}$, $\rm I_{\emph{k}}^{UL}(\emph{t})=\sum\limits_{\emph{m}\in N^{+},\emph{m}\neq \emph{k}}p_{T-RR{U_\emph{m}}}(\emph{t})\cdot
h_{\emph{m}_\emph{k}}^{RRU-RRU}(\emph{t})\cdot
r_{\emph{m}_\emph{k}}^{RRU-RRU}{(\emph{t})^{-\alpha}}$ is the aggregated interference from adjacent BSs except for the desired BS $\rm {BS}_\emph{k}$.

\subsection{Weighted Energy Efficiency}
Generally speaking, the BS in particular RRU and the UE have different power sources and have different requirements on energy efficiency. For example, a smart phone (i.e., UE) uses a battery to store the power and requires to be recharged when the battery is used up. The RRU can have more power supply from the wired cable. In this paper, we consider a weighted sum of the energy efficiency for both UE and BS defined as follows:
\begin{equation}
\label{3}
\rm U_{B{S_\emph{k}}}(\emph{t}) = \frac{\sum\limits_{\emph{i}_{DL}}EE_{\emph{k}_{\emph{i}_{DL}}}^{DL}(\emph{t})}{N_{DL}}+
\beta\cdot\frac{\sum\limits_{\emph{i}_{UL}}EE_{\emph{k}_{\emph{i}_{UL}}}^{UL}(\emph{t})}{N_{UL}},
\end{equation}
where $\beta$ is the weighted factor for uplinks. $\rm EE_{\emph{k}_{\emph{i}_{DL}}}^{DL}(\emph{t})$ is the energy efficiency of downlinks between the desired BS $\rm {BS}_\emph{k}$ and the UE $\rm {UE}_{\emph{i}_{DL}}$ at the time slot $t$,  $\rm N_{DL}$ is the active number of UEs in downlinks, $\rm EE_{\emph{k}_{\emph{i}_{UL}}}^{UL}(\emph{t})$ is the energy efficiency of uplinks between the desired UE $\rm {UE}_{\emph{i}_{UL}}$ and the BS $\rm {BS}_\emph{k}$ at the time slot $t$, $\rm N_{UL}$ is the active number of UEs in uplinks.

Without loss of generality,  the total power consumption of the BS $\rm {BS}_\emph{k}$ is divided into the RRU transmission power $\rm p_{T-RR{U_\emph{k}}}(\emph{t})$ and the static power $\rm p_{static}$. The static power $\rm p_{static}$ is fixed for all BSs. To simplify derivations,  bandwidths of downlinks and uplinks are normalized as 1. Hence,  we define the energy efficiency of the BS in time slot $t$ as the number of data bits that can be sent by the BS per unit of power consumption, which is denoted by,
\begin{equation}
\label{4}
\rm EE_{{\emph{k}_\emph{i}}}^{DL}(\emph{t}) = \frac{\ln(1+SINR_{\emph{k}_\emph{i}}^{DL}(\emph{t}))}
{p_{T-RR{U_\emph{k}}}(\emph{t})+p_{static}}.
\end{equation}
Similarly, we define the energy efficiency of the UE as the number of bits of data sent per unit of consumed power given by
\begin{equation}
\label{5}
\rm EE_{\emph{k}_\emph{i}}^{UL}(\emph{t}) = \frac{\ln(1+SINR_{\emph{k}_\emph{i}}^{UL}(\emph{t}))}
{p_{T-UE}(\emph{t})+p_{UE-static}},
\end{equation}
where $\rm p_{UE-static}$ is the static power at UEs and is assumed to be fixed for all UEs, $\rm p_{T-UE}(\emph{t})$ is the transmission power at UEs. The total power consumption of the UE includes the static power and the transmission power at UEs.
The self-interference is the interference from the transmitting antenna to the receiving antenna at the same node. Since the size of a node generally is not very large, the link distance is relatively short. Therefore, the self-interference could be considered as fixed.
\subsection{Energy Efficiency Formulation of BSs}
Based on the system model in Fig. 1(a),  the RRUs are supported by batteries. In this case,  the RRU transmission power of the BS $\rm {BS}_\emph{k}$ is ranged by $\rm p_{T-RR{U_\emph{k}}}(\emph{t})\in [0, p_{T-RR{U_{max}}}]$, where $\rm p_{T-RR{U_{max}}}$ is the maximum transmission power supported by the battery. Since the wireless transmission is supported by the battery energy of RRUs, the battery level of RRUs depends on the transmission power and varies with the transmission time. The battery capacity of RRU is given by $\rm E_{max}$. Let $e_k(t)$ be the battery level of the $k$th RRU.
at the time $t$. Hence, the value of $\rm e_\emph{k}(\emph{t})$ is ranged as $\rm 0\leq e_\emph{k}(\emph{t})\leq E_{max}$. Without loss of generality, the batteries of RRUs are assumed to be wirelessly charged by BBU after a period $T$. The state equation of battery volume at the RRU is defined as follows:

\vspace{0.1cm}
\textbf{Definition 1 (State equation):}
We define the state of battery for BS $k$ as the battery level in each time, denoted as $e_k(t)$ of RRU $\rm e_\emph{k}(\emph{t})$. The change of the residual battery volume of RRU is expressed by the following state equation:
\begin{equation}
\label{6}
\rm de_\emph{k}(\emph{t}) = -p_{T-RR{U_\emph{k}}}(\emph{t})d\emph{t}, 0\leq \emph{t}\leq \emph{T}
\end{equation}

In this paper the transmission power of RRUs is assumed to be adaptively adjusted by the residual battery volume of RRUs. Therefore,  the RRU transmission power of the BS $\rm {BS}_\emph{k}$ $\rm p_{T-RR{U_\emph{k}}}(\emph{t})$ is replaced by $\rm p_{T-RR{U_\emph{k}}}(\emph{t}, e_\emph{k}(\emph{t}))$, which is simply denoted as $\rm p_{T-RR{U_\emph{k}}}(\emph{t}, e_\emph{k})$. When the $\rm p_{T-RR{U_\emph{k}}}(\emph{t})$ is replaced by $\rm p_{T-RR{U_\emph{k}}}(\emph{t}, e_\emph{k})$ in $(4)$ and $(5)$,  the weighted energy efficiency of a cell associated with the BS $\rm {BS}_\emph{k}$  is denoted as $\rm U_{B{S_\emph{k}}}(p_{T-RR{U_\emph{k}}}(\emph{t}, e_\emph{k}))$.

Considering the wireless power transfer technology adopted for RRUs, the battery volume of RRU is recharged after a time interval $T$. Hence, the total power consumed by an RRU is limited in the time interval $T$. To maximize the energy efficiency of BSs in the time interval $T$, not only the weighted energy efficiency of cells but also the residual battery volume of RRU at BSs need to be considered for optimizing the transmission power. Based on (5) and (6), the maximum energy efficiency of cell with the BS $\rm {BS}_\emph{k}$ in the time interval $T$ is formulated as
\begin{equation}
\label{7}
\rm \max\limits_{p_{T-RR{U_\emph{k}}}(0\rightarrow \emph{T})}E[{\int_0^\emph{T}}U_{BS_\emph{k}}(p_{T-RR{U_\emph{k}}}(\emph{t}, e_\emph{k}))d\emph{t}]
\end{equation}

$$
\begin{aligned}
\mathrm{s.t.}\quad&\rm 0\leq p_{T-RR{U_\emph{k}}}(\emph{t}, \emph{e}_\emph{k})\leq p_{T-RR{U_{max}}},  \\
\quad&\rm d\emph{e}_\emph{k}(\emph{t})=-p_{T-RR{U_\emph{k}}}(\emph{t}, \emph{e}_\emph{k})d\emph{t}, 0\leq \emph{t}\leq \emph{T}.\\
\end{aligned}
$$
where (7) is the optimal function maximizing the average energy efficiency of BSs in the time interval $T$, $\rm p_{T-RR{U_\emph{k}}}(0\rightarrow \emph{T})$ is the RRU transmission power strategy of the BS $\rm BS_\emph{k}$ in the time interval $T$.
Based on the stochastic optimal control theory in \cite{Basar33}, the optimal RRU transmission power strategy $\rm p_{T-RR{U_\emph{k}}}^*(0\rightarrow \emph{T})$ of the BS $\rm BS_\emph{k}$ is given to achieve the result of (7). The expression of $\rm p_{T-RR{U_\emph{k}}}^*(0\rightarrow \emph{T})$ is expressed as follows:
\begin{equation}
\label{8}
\rm \arg\max\limits_{p_{T-RR{U_\emph{k}}}(0\rightarrow \emph{T})}E[{\int_0^\emph{T}}U_{BS_\emph{k}}(p_{T-RR{U_\emph{k}}}(\emph{t}, e_\emph{k}))d\emph{t}].
\end{equation}

To illustrate the maximum of energy efficiency of cells with different RRU transmission power strategies from the time $t$ to $T$, the value function $\rm V_\emph{k}(\emph{t}, e_\emph{k})$ of the BS $\rm BS_\emph{k}$ is defined, which is given as
\begin{equation}
\label{9}
\rm \max\limits_{p_{T-RR{U_\emph{k}}}(\emph{t}\rightarrow \emph{T})}E[{\int_\emph{t}^\emph{T}}U_{BS_\emph{k}}(p_{T-RR{U_\emph{k}}}(\tau, e_\emph{k}))d\tau],\emph{t}\in[0, \emph{T}].
\end{equation}

Based on the Bellman principle of optimality in \cite{Bell34}, for the optimal strategy of RRU transmission power, the RRU transmission power strategy is always the optimal strategy from any start time to the end time. Therefore,  the optimal strategy of RRU transmission power is defined as follows:

\textbf{Definition 2 (The optimal strategy of RRU transmission power):}
 When $\rm p_{T-RR{U_\emph{k}}}^*(\emph{t}\rightarrow \emph{T})$ is the optimal strategy of RRU transmission power at the BS $\rm BS_\emph{k}$,  for any time slot $t\in[0, T]$, the following equation is always satisfied
\begin{equation}
\label{10}
\rm E[{\int_\emph{t}^\emph{T}}U_{BS_\emph{k}}(p_{T-RR{U_\emph{k}}}^*
(\tau, e_\emph{k}))d\tau] = V_\emph{k}(\emph{t}, e_\emph{k}),
\end{equation}
where $\rm p_{T-RR{U_\emph{k}}}^*(\tau, e_\emph{k})$ is the RRU transmission power of the BS $\rm BS_\emph{k}$ at the time slot $\tau$  when the optimal strategy of RRU transmission power $\rm p_{T-RR{U_\emph{k}}}^*(\emph{t}\rightarrow \emph{T})$  is adopted for the BS $\rm BS_\emph{k}$. The result of $(10)$ implies that the (7) always achieves the maximum in the time internal $T$ when the optimal RRU transmission power strategy $\rm p_{T-RR{U_\emph{k}}}^*(\emph{t}\rightarrow \emph{T})$  is adopted at the BS $\rm BS_\emph{k}$.

\section{Network Energy Efficiency Optimization}
\subsection{Differential Game}
The maximum energy efficiency in (7) is formulated for the single cell. However,  the optimal RRU transmission power strategy is influenced between each other by the interference from adjacent RRUs in 5G cellular networks. Hence, the optimal RRU transmission power strategy is not independent from each other. Therefore,  we use the differential game theory \cite{Han35} to optimize the RRU transmission power strategy in 5G cellular networks. In this paper $\varphi_G$ is defined for the differential game among the RRU transmission power strategy in 5G cellular networks.

\textbf{Definition 3 (Differential game of RRU transmission power strategy $\varphi_G$):}
Every BS is assumed as a player for a cellular network. The utility function of player $k$ is expressed as
\begin{equation}
\label{11}
\rm UF_\emph{k} = E[{\int_0^\emph{T}}U_{BS_\emph{k}}(p_{T-RR{U_\emph{k}}}(\emph{t}, e_\emph{k}))d\emph{t}],
\end{equation}
with the object function:
\[\rm \max\limits_{p_{T-RR{U_\emph{k}}}(0\rightarrow \emph{T})}UF_\emph{k}\]

$$
\begin{aligned}
\mathrm{s.t.}\quad&\rm 0\leq p_{T-RR{U_\emph{k}}}(\emph{t}, \emph{e}_\emph{k})\leq p_{T-RR{U_{max}}},  \\
\quad&\rm d\emph{e}_\emph{k}(\emph{t})=-p_{T-RR{U_\emph{k}}}(\emph{t}, \emph{e}_\emph{k})d\emph{t}, 0\leq \emph{t}\leq \emph{T}.\\
\end{aligned}
$$

Every BS in the cellular network , {\em i.e.,} every player in the differential game, always trys to achieve the maximum utility function of player by adjusting the transmission power of RRU. In this case,  the RRU transmission power strategy of BSs is adaptively adjusted by a game method. In the end, the RRU transmission power strategy finally selected by every BS will approach a stable state and is the optimal for cellular networks. When each BS chooses the optimal RRU transmission power strategy to maximize its own utility, the obtained solution for the overall optimization problem achieves a Nash equilibrium. The Nash equilibrium of differential game of RRU transmission power strategy is defined as:

\textbf{Definition 4 (Nash Equilibrium of the Differential game of RRU transmission power strategy $\varphi_G$):}
If and only if the following equation, {\em i.e.,} (12) is satisfied, the strategy set of all players $\rm p^* = [p_{T-RR{U_1}}^*(0\rightarrow \emph{T}), ..., p_{T-RR{U_\emph{k}}}^*(0\rightarrow \emph{T}), ...]$ is the Nash equilibrium of differential game of RRU transmission power strategy,
\begin{equation}
\label{12}
\rm p_{T-RR{U_\emph{k}}}^*(0\rightarrow \emph{T})= \arg\max\limits_{p_{T-RR{U_\emph{k}}}(0\rightarrow \emph{T})}UF_\emph{k},
\end{equation}
subject to:
\[\rm de_\emph{k}(\emph{t}) = -p_{T-RR{U_\emph{k}}}(\emph{t}, e_\emph{k})d\emph{t}, 0\leq \emph{t}\leq \emph{T}.\]

Assumed that $\rm p_{-\emph{k}}^*$  is the strategy set of players without the player $k$,{\em i.e.,} the BS $\rm BS_\emph{k}$. Based on the Nash equilibrium in (12), the BS $\rm BS_\emph{k}$ adopting the strategy $\rm p_{T-RR{U_\emph{k}}}^*(0\rightarrow \emph{T})$ can achieve the maximum utility function when other BSs adopt the strategy set $\rm p_{-\emph{k}}^*$. At Nash equilibrium, the value of utility function at a BS will be less than the maximum of utility function when the BS changes the current RRU transmission power strategy. Therefore,  all BSs would not like to change the current RRU transmission power strategy based on the Nash equilibrium. The RRU transmission power of all BSs approaches to a stable state.

To derive the solution of Nash equilibrium $\varphi_G$ for the RRU transmission power strategy of BSs,  the existence of the Nash equilibrium $\varphi_G$ should be firstly proved.
\begin{thm}
The Nash equilibrium $\varphi_G$ exists for the proposed RRU transmission power strategy.
\end{thm}

Based on the results in \cite{Soner36}, the Nash equilibrium is existed in the differential game if the Hamilton-Jacobi-Bellman (HJB) equations of all players can be solved. The HJB equation of BS $\rm BS_\emph{k}$ is expressed as (13),
\begin{figure*}[!t]
\begin{equation}
\label{13}
\rm \frac{\partial V_\emph{k}(\emph{t}, e_\emph{k})}{\partial \emph{t}}+\max\limits_{p_{T-RR{U_\emph{k}}}(\emph{t}, e_\emph{k})}[U_{BS_\emph{k}}(p_{T-RR{U_\emph{k}}}(\emph{t}, e_\emph{k}))-p_{T-RR{U_\emph{k}}}(\emph{t}, e_\emph{k})\cdot\frac{\partial V_\emph{k}(\emph{t}, e_\emph{k})}{\partial e}] = 0
\end{equation}
\begin{equation}
\label{14}
\rm H(e_\emph{k}, \frac{\partial V_\emph{k}(\emph{t}, e_\emph{k})}{\partial e})
\rm = \max\limits_{p_{T-RR{U_\emph{k}}}(\emph{t}, e_\emph{k})}(\frac{\sum\limits_{\emph{i}_{DL}}\frac{\ln(1+SINR_{k_i}^{\rm DL}(\emph{t}))}{p_{T-RR{U_\emph{k}}}(\emph{t}, e_\emph{k})+p_{static}}}{N_{DL}}
\rm +\beta\cdot \frac{\sum\limits_{\emph{i}_{UL}}\frac{\ln(1+SINR_{\emph{k}_\emph{i}}^{UL}(\emph{t}))}{p_{T-UE}+p_{UE-static}}}{N_{UL}}
\rm -p_{T-RR{U_\emph{k}}}(\emph{t}, e_\emph{k})\cdot\frac{\partial V_\emph{k}(\emph{t}, e_\emph{k})}{\partial e})\hfill
\end{equation}
\hrulefill
\end{figure*}
To easily analyze the characteristic of $(13)$, we set $\rm H(e_\emph{k}, \frac{\partial V_\emph{k}(t, e_\emph{k})}{\partial e})=\max\limits_{p_{T-RR{U_\emph{k}}}(\emph{t},e_\emph{k})}[U_{BS_\emph{k}}(p_{T-RR{U_\emph{k}}}(\emph{t},e_\emph{k}))-p_{T-RR{U_\emph{k}}}(\emph{t}, e_\emph{k})\cdot\frac{\partial {V_\emph{k}}(\emph{t}, e_\emph{k})}{\partial e}]$ as the Hamiltonian function. Hence,  $(13)$  can be solved only if the Hamiltonian function can be solved. Based on the results in \cite{Evans37}, the condition that the Hamiltonian function can be solved is that the Hamiltonian function must be a smooth function, {\em i.e., }the derivative of the Hamiltonian function on $\rm p_{T-RR{U_\emph{k}}}(\emph{t}, e_\emph{k})$ exists in the available domain. Substitute $(1)$, $(2)$, $(3)$, $(4)$ and $(5)$ into the Hamiltonian function,  the Hamiltonian function can be obtained as (14).

In this paper $\rm p_{T-RR{U_\emph{k}}}(\emph{t}, e_\emph{k})$  is configured as the differential variable for the Hamiltonian function. The term of $\rm \frac{\partial V_\emph{k}(\emph{t}, e_\emph{k})}{\partial e}$ is independent on the RRU transmission power strategy of BS,  so the derivative of $\rm \frac{\partial V_\emph{k}(\emph{t}, e_\emph{k})}{\partial e}$ on $\rm p_{T-RR{U_\emph{k}}}(\emph{t}, e_\emph{k})$ is zero. $\rm N_{DL}$ and $\rm N_{UL}$ are considered as constants for the differential variable $\rm p_{T-RR{U_\emph{k}}}(\emph{t},  e_\emph{k})$. $\rm I_{\emph{k}}^{DL}(\emph{t})$ and $\rm I_{\emph{k}}^{UL}(\emph{t})$ are independent on $\rm p_{T-RR{U_\emph{k}}}(\emph{t}, e_\emph{k})$. As a consequence, the Hamiltonian function is an elementary function when $\rm p_{static}> 0$. Therefore, the derivative of the Hamiltonian function on $\rm p_{T-RR{U_\emph{k}}}(\emph{t}, e_\emph{k})$  exists in the available domain,  {\em i.e.,} the Hamiltonian function is a smooth function. Furthermore,  the Hamiltonian function can be solved and then $(13)$  can also be solved,  {\em i.e., }the Nash equilibrium can be solved by the HJB equations. In the end,  the existence of the Nash equilibrium $\varphi_G$  is proved for the RRU transmission power strategy of BSs.

\subsection{Mean Field Game}
Generally the solution of Nash equilibrium can be derived by solving a set of partial differential equations. Moreover,  the number of partial differential equations used to solve the Nash equilibrium is equal to the number of BSs in the cellular network. For a real cellular network,  the number of BSs is very large. Hence,  the number of partial differential equations is too large to solve the Nash equilibrium for a real cellular network. On the other hand,  the distribution function of the residual battery volume of RRU at BSs is approximated as a continuous function when the number of BSs is large enough in the cellular network. In this case,  the impact of the RRU transmission power strategy from the single BS on the RRU transmission power strategy of another BS can be ignored and then only the impact of the cellular network on the RRU transmission power strategy of the single BS needs to be considered. As a consequence, the differential game can be replaced by the mean field game to optimize the energy efficiency of cellular networks.

To describe the probability density function (PDF) of residual battery volume of RRUs, the mean field of network energy efficiency is defined as
\begin{equation}
\label{15}
\rm m(\emph{t}, \emph{e}) = \lim\limits_{N\rightarrow\infty}\frac{1}{N}\sum\limits_{\emph{k}=1}^{N}I_{\{e_\emph{k}(\emph{t})=e\}},
\end{equation}
where $\rm I_{\{e_\emph{k}(\emph{t})=e\}}$ is equal to 1 when the residual battery volume of RRU at the BS $\rm BS_\emph{k}$ is equal to $\rm e$, otherwise $\rm I_{\{e_\emph{k}(\emph{t})=e\}}$ is equal to 0. $\rm N$ is the number of BSs. $\rm m(\emph{t}, \emph{e})$ can be used to describe the distribution of the residual battery volume of RRUs on the time slot $t$. When the number of BSs is large, $\rm m(\emph{t}, \emph{e})$ can be assumed as a continuous function.

Considering the mean field definition in $(15)$, $\rm I_{mean-RR{U_\emph{k}}}(\emph{t})$ is denoted as the interference aggregated at the BS $\rm BS_\emph{k}$ and $\rm I_{mean-U{E_{\emph{i}_{DL}}}}(\emph{t})$ is denoted as the interference aggregated at the user $\rm UE_{\emph{i}_{DL}}$ which is associated with the BS $\rm BS_\emph{k}$. The detail expressions of $\rm I_{mean-RR{U_\emph{k}}}(\emph{t})$ and $\rm I_{mean-U{E_{\emph{i}_{DL}}}}(\emph{t})$ are derived as
\begin{equation}
\label{16}
\begin{aligned}
&\rm I_{mean-RR{U_\emph{k}}}(\emph{t}) \hfill \\
&\rm = E[\sum\limits_{\emph{m}\neq \emph{k}}p_{T-RR{U_\emph{m}}}(\emph{t}, e_\emph{k})h_{\emph{m}_\emph{k}}^{RRU-RRU}(\emph{t}) r_{\emph{m}_\emph{k}}^{RRU-RRU}(\emph{t})^{-\alpha}] \hfill \\
&\rm = E[p_{T-RR{U_\emph{m}}}(\emph{t}, e_\emph{k})]E[h_{\emph{m}_\emph{k}}^{RRU-RRU}(\emph{t})]E[\sum\limits_{\emph{m}\neq \emph{k}}r_{\emph{m}_\emph{k}}^{RRU-RRU}(\emph{t})^{-\alpha}], \hfill
\end{aligned}
\end{equation}
\begin{equation}
\label{17}
\begin{aligned}
&\rm I_{mean-U{E_{\emph{i}_{DL}}}}(\emph{t}) \hfill \\
&=\rm E[\sum\limits_{\emph{m}\neq \emph{k}}p_{T-RR{U_\emph{m}}}(\emph{t}, e_\emph{k}) h_{\emph{m}_{\emph{i}_{DL}}}^{RRU-UE}(\emph{t}) r_{\emph{m}_{\emph{i}_{DL}}}^{RRU-UE}(\emph{t})^{-\alpha}] \hfill \\
&\rm = E[p_{T-RR{U_\emph{m}}}(\emph{t}, e_\emph{k})] E[h_{\emph{m}_{\emph{i}_{DL}}}^{RRU-UE}(\emph{t})]\cdot E[\sum\limits_{\emph{m}\neq \emph{k}}r_{\emph{m}_{\emph{i}_{DL}}}^{RRU-UE}(\emph{t})^{-\alpha}],
\end{aligned}
\end{equation}
where $\rm E[p_{T-RR{U_\emph{m}}}(\emph{t}, e_\emph{k})]$  is the expectation of RRU transmission power from interfering BSs on the time slot $t$. Based on the system model and properties of Poisson distribution \cite{Andrews39}, the distributions of $\rm r_{\emph{m}_\emph{k}}^{RRU-RRU}(\emph{t})$ and $\rm r_{\emph{m}_{\emph{i}_{DL}}}^{RRU-UE}(t)$ are the same. When $\rm r_{\emph{m}_\emph{k}}^{RRU-RRU}(\emph{t})< 1$ or $\rm r_{\emph{m}_{\emph{i}_{DL}}}^{RRU-UE}(\emph{t})< 1$, the path loss fading is configured as $\alpha = 1$. Moreover, wireless channels of $\rm h_{\emph{m}_\emph{k}}^{RRU-RRU}(\emph{t})$ and $\rm h_{\emph{m}_{\emph{i}_{DL}}}^{RRU-UE}(\emph{t})$ are assumed to be governed by the exponent distribution with the mean as one \cite{Sema44}. Based on the Campbell's theorem \cite{Haen38} and the properties of cumulative distribution function (CDF) in Poisson distribution \cite{Andrews39}, $\rm I_{mean-RR{U_\emph{k}}}(\emph{t})$ and $\rm I_{mean-U{E_{\emph{i}_{DL}}}}(\emph{t})$ are further
derived as
\begin{equation}
\label{18}
\begin{aligned}
&\rm I_{mean-RR{U_\emph{k}}}(\emph{t})\\
&\rm = E[p_{T-RR{U_\emph{m}}}(\emph{t}, e_\emph{k})] E[\sum\limits_{\emph{m}\neq \emph{k}}r_{\emph{m}_\emph{k}}^{RRU-RRU}(\emph{t})^{-\alpha}] \hfill \\
& \rm =E[p_{T-RR{U_\emph{m}}}(\emph{t},e_\emph{k})]{\int_{S\backslash\{\emph{k}\}}}r_{\emph{m}_\emph{k}}^{RRU-RRU}(t)^{-\alpha}d(S)\hfill \\
&\rm = E[p_{T-RR{U_\emph{m}}}(\emph{t}, e_\emph{k})]\cdot 2\pi\lambda_{BS}\cdot(\frac{1}{2}+\frac{1}{\alpha-2})\hfill
\end{aligned}
\end{equation}
\begin{equation}
\begin{aligned}
&\rm I_{mean-U{E_{\emph{i}_{DL}}}}(\emph{t})\\
&\rm= E[p_{T-RR{U_\emph{m}}}(\emph{t}, e_\emph{k})] E[\sum\limits_{\emph{m}\neq \emph{k}}r_{\emph{m}_{\emph{i}_{DL}}}^{RRU-UE}(\emph{t})^{-\alpha}] \hfill \\
&\rm = E[p_{T-RR{U_\emph{m}}}(\emph{t}, e_\emph{k})]\cdot{\int_{S\backslash\{\emph{k}\}}}r_{\emph{m}_{\emph{i}_{DL}}}^{RRU-UE}(\emph{t})^{-\alpha}d(S)\hfill \\
&\rm = E[p_{T-RR{U_\emph{m}}}(\emph{t}, e_\emph{k})] 2\pi\lambda_{BS}\cdot(\frac{1}{2}+\frac{1}{\alpha-2})\hfill
\end{aligned}
\end{equation}
where $\rm S\backslash\{\emph{k}\}$ denotes the set of BSs in a cellular network without the BS $\rm BS_\emph{k}$. $\rm p_T(\emph{t}, \emph{e})$ is configured as the average RRU transmission power with the residual battery volume $e$ on the time slot $t$, {\em i.e.,}
$\rm p_T(\emph{t}, \emph{e}) = \lim\limits_{N\rightarrow\infty}{\frac{1}{N}\sum\limits_{\emph{k}=1}^{N}p_{T-RR{U_\emph{k}}}(\emph{t},\emph{ e}_\emph{k})\cdot I_{\{\emph{e}_\emph{k}(\emph{t})=\emph{e}\}}}$. Hence, the expectation of the RRU transmission power in the cellular network can be derived as
\begin{equation}
\rm E[p_{T-RR{U_\emph{m}}}(\emph{t}, \emph{e}_\emph{k})] = \int_{\emph{e}\in\varepsilon}p_T(\emph{t}, \emph{e})\cdot m(\emph{t}, \emph{e})d\emph{e}.
\end{equation}
Furthermore,  the average interference in the mean field of network energy efficiency is derived as
\begin{equation}
\begin{aligned}
&\rm I_{mean-RR{U_\emph{k}}}(\emph{t}) \\
&\rm= I_{mean-U{E_{\emph{i}_{DL}}}}(\emph{t})\\
&\rm = 2\pi\lambda_{BS}\cdot(\frac{1}{2}+\frac{1}{\alpha-2})\cdot\int_{\emph{e}\in\varepsilon}p_T(\emph{t}, \emph{e})\cdot m(\emph{t}, \emph{e})d\emph{e}.
\end{aligned}
\end{equation}

Based on the results in \cite{Al43}, the interference fluctuation in the small scale can be smoothed when a large number of BSs exists. Hence, $\rm I_{mean-RR{U_\emph{k}}}(\emph{t})$ and $\rm I_{mean-U{E_{\emph{i}_{DL}}}}(\emph{t})$ are considered as stationary values at the time slot $t$. $\rm EE_{\emph{k}}^{DL-mean}(\emph{t})$ denotes the energy efficiency of downlinks in the mean field of network energy efficiency and the calculation of $\rm EE_{\emph{k}}^{DL-mean}(\emph{t})$ needs to average the energy efficiency of all users accessing with the BS. $\rm EE_{\emph{k}}^{DL-mean}(\emph{t})$ is derived as (22a), (22b), (22c).
\begin{figure*}[!t]
\[\begin{gathered}
\rm EE_{\emph{k}}^{DL-mean}(\emph{t}) \rm= {\frac{\sum\limits_{\emph{i}_{DL}}EE_{\emph{k}_{\emph{i}_{DL}}}^{DL}(\emph{t})}{N_{DL}}}
\rm =E[EE_{\emph{k}_{\emph{i}_{DL}}}^{DL}(\emph{t})]
\rm={\frac{E[\ln(1+SINR_{\emph{k}_{\emph{i}_{DL}}}^{DL}(\emph{t}))]}
{p_{T-RR{U_\emph{k}}}(\emph{t}, \emph{e}_\emph{k})+p_{static}}}\end{gathered},\qquad\qquad\qquad\qquad\qquad\quad\quad \tag{22a}\]
\[\begin{gathered}
\rm E[\ln(1+SINR_{\emph{k}_{\emph{i}_{DL}}}^{DL}(\emph{t}))]
= \int\limits_{\omega>0}Pr[SINR_{\emph{k}_{\emph{i}_{DL}}}^{DL}(\emph{t})>\emph{e}^\omega-1]d\omega\end{gathered},\qquad\qquad\qquad\qquad\qquad
\qquad\qquad\qquad\quad\quad\,\,\tag{22b}\]
\[\begin{gathered}
\rm Pr[SINR_{\emph{k}_{\emph{i}_{DL}}}^{DL}(\emph{t})>\emph{e}^\omega-1] \hfill \\
\rm \quad = E_{r_{\emph{k}_{\emph{i}_{DL}}}^{RRU-UE}(\emph{t})}
[Pr[SINR_{\emph{k}_{\emph{i}_{DL}}}^{DL}(\emph{t})>\emph{e}^\omega-1|r_{\emph{k}_{\emph{i}_{DL}}}^{RRU-UE}(\emph{t})]]\hfill \\
\rm\quad =\int_{r_{\emph{k}_{\emph{i}_{DL}}}^{RRU-UE}(\emph{t})>0}Pr[SINR_{\emph{k}_{\emph{i}_{DL}}}^{DL}(\emph{t})>e^\omega-1|
r_{\emph{k}_{\emph{i}_{DL}}}^{RRU-UE}(\emph{t})]f(r_{\emph{k}_{\emph{i}_{DL}}}^{RRU-UE}(\emph{t})) \hfill \\
\rm\quad =\int_{r_{\emph{k}_{\emph{i}_{DL}}}^{RRU-UE}(\emph{t})>0}Pr[h_{\emph{k}_{\emph{i}_{DL}}}^{RRU-UE}(\emph{t})
>{\frac{(\emph{e}^\omega-1)\cdot(I_{mean-U{E_{\emph{i}_{DL}}}}(t)+N_0)\cdot r_{\emph{k}_{\emph{i}_{DL}}}^{RRU-UE}(\emph{t})^\alpha}
{p_{T-RR{U_\emph{k}}}(\emph{t}, \emph{e}_\emph{k})}}|r_{\emph{k}_{\emph{i}_{DL}}}^{RRU-UE}(\emph{t})]\hfill \\
\qquad \rm f(r_{\emph{k}_{\emph{i}_{DL}}}^{RRU-UE}(\emph{t}))dr_{\emph{k}_{\emph{i}_{DL}}}^{RRU-UE}(\emph{t}) \hfill \\
\rm \quad =\int_{r_{\emph{k}_{\emph{i}_{DL}}}^{RRU-UE}(\emph{t})>0}\exp(-{\frac{(\emph{e}^\omega-1)\cdot(I_{mean-U{E_{\emph{i}_{DL}}}}(\emph{t})+N_0)\cdot r_{\emph{k}_{\emph{i}_{DL}}}^{RRU-UE}(\emph{t})^\alpha}{p_{T-RR{U_\emph{k}}}(\emph{t}, \emph{e}_\emph{k})}})
\rm f(r_{\emph{k}_{\emph{i}_{DL}}}^{RRU-UE}(\emph{t}))dr_{\emph{k}_{\emph{i}_{DL}}}^{RRU-UE}(\emph{t}) \hfill
\end{gathered},\tag{22c}\]

\[\begin{gathered}
\rm \Phi_1=\int\limits_{\omega>0}\int\limits_{r_{\emph{k}_{\emph{i}_{DL}}}^{RRU-UE}(\emph{t})>0}
\exp(-{\frac{(\emph{e}^\omega-1)\cdot(I_{mean-U{E_{\emph{i}_{DL}}}}(\emph{t})+N_0)\cdot r_{\emph{k}_{\emph{i}_{DL}}}^{RRU-UE}(\emph{t})^\alpha}
{p_{T-RR{U_\emph{k}}}(\emph{t}, \emph{e}_\emph{k})}})2\pi\lambda_{BS}r_{\emph{k}_{\emph{i}_{DL}}}^{RRU-UE}(\emph{t})\\
\rm\quad\quad\quad \emph{e}^{-\lambda_{BS}\pi(r_{\emph{k}_{\emph{i}_{DL}}}^{RRU-UE}(\emph{t}))^2}dr_{\emph{k}_{\emph{i}_{DL}}}^{RRU-UE}(\emph{t})d\omega \hfill
\end{gathered}.\quad\quad\tag{22d}
\]
\hrulefill
\vspace*{4pt}
\end{figure*}
$\rm f(r_{\emph{k}_{\emph{i}_{DL}}}^{RRU-UE}(\emph{t})) = 2\pi\lambda_{BS}r_{\emph{k}_{\emph{i}_{DL}}}^{RRU-UE}(\emph{t})
\emph{e}^{-\lambda_{BS}\pi(r_{\emph{k}_{\emph{i}_{DL}}}^{RRU-UE}(\emph{t}))^2}$ is the PDF between the UE $\rm UE_{\emph{i}_{DL}}$ and the closest BS. In the end,  the energy efficiency of downlinks in the mean field of network energy efficiency is expressed as (23). And the $\Phi_1$ is illustrated in (22d).
\[\rm EE_{\emph{k}}^{DL-mean}(\emph{t}) = {\frac{\Phi_1}
{p_{T-RR{U_\emph{k}}}(\emph{t}, \emph{e}_\emph{k})+p_{static}}}, \tag{23}\]

$\rm EE_{\emph{k}}^{UL-mean}(\emph{t})$ denotes the energy efficiency of uplinks in the mean field of network energy efficiency. Based on the derivation process in $(23)$, $\rm EE_{\emph{k}}^{UL-mean}(\emph{t})$ is similarly derived as (24a), (24b), (24c).
\begin{figure*}[!t]
\[\begin{gathered}\rm EE_{\emph{k}}^{UL-mean}(\emph{t}) = {\frac{\sum\limits_{\emph{i}_{UL}}EE_{\emph{k}_{\emph{i}_{UL}}}^{UL}(\emph{t})}{N_{UL}}}
\rm =E[EE_{\emph{k}_{\emph{i}_{UL}}}^{UL}(\emph{t})]={\frac{E[\ln(1+SINR_{\emph{k}_{\emph{i}_{UL}}}^{UL}(\emph{t}))]}
{p_{T-RR{U_\emph{k}}}(\emph{t}, \emph{e}_\emph{k})+p_{static}}}\end{gathered},\qquad\qquad\qquad\quad\quad\quad\qquad\,\,\,\,\,
\tag{24a}\]
\[\begin{gathered}
\rm E[\ln(1+SINR_{\emph{k}_{\emph{i}_{UL}}}^{UL}(\emph{t}))]
\rm = \int\limits_{\omega>0}Pr[SINR_{\emph{k}_{\emph{i}_{UL}}}^{UL}(\emph{t})>\emph{e}^\omega-1]d\omega \hfill
\end{gathered}, \qquad\qquad\qquad\qquad\qquad\qquad\quad\quad\qquad\quad\quad\tag{24b}\]
\[\begin{gathered}
\rm Pr[SINR_{\emph{k}_{\emph{i}_{UL}}}^{UL}(\emph{t})>\emph{e}^\omega-1] \hfill \\
\rm\quad = E_{r_{\emph{k}_{\emph{i}_{UL}}}^{RRU-UE}(\emph{t})}
\rm [Pr[SINR_{\emph{k}_{\emph{i}_{UL}}}^{UL}(\emph{t})>\emph{e}^\omega-1|r_{\emph{k}_{\emph{i}_{UL}}}^{RRU-UE}(\emph{t})]] \hfill \\
\rm\quad =\int\limits_{r_{\emph{k}_{\emph{i}_{UL}}}^{RRU-UE}(\emph{t})>0}Pr[SINR_{\emph{k}_{\emph{i}_{UL}}}^{UL}(\emph{t})>\emph{e}^\omega-1|
r_{\emph{k}_{\emph{i}_{UL}}}^{RRU-UE}(\emph{t})]f(r_{\emph{k}_{\emph{i}_{UL}}}^{RRU-UE}(\emph{t}))\hfill \\
\rm\quad =\int\limits_{r_{\emph{k}_{\emph{i}_{UL}}}^{RRU-UE}(\emph{t})>0}\exp(-{\frac{(\emph{e}^\omega-1)\cdot(I_{mean-RR{U_\emph{k}}}(\emph{t})+p_{T-RR{U_\emph{k}}}(\emph{t})\cdot h_{self}(\emph{t})+N_0)\cdot r_{\emph{k}_{\emph{i}_{UL}}}^{RRU-UE}(\emph{t})^\alpha}{p_{T-RR{U_\emph{k}}}(\emph{t}, \emph{e}_\emph{k})}}) \hfill \\
\rm \quad\quad f(r_{\emph{k}_{\emph{i}_{UL}}}^{RRU-UE}(\emph{t}))dr_{\emph{k}_{\emph{i}_{UL}}}^{RRU-UE}(\emph{t}) \hfill
\end{gathered},\quad\quad\tag{24c}\]
\[\begin{gathered}
\rm \Phi_2=\int\limits_{\omega>0}\int\limits_{r_{\emph{k}_{\emph{i}_{UL}}}^{RRU-UE}(\emph{t})>0}
\rm 2\pi\lambda_{BS}r_{\emph{k}_{\emph{i}_{UL}}}^{RRU-UE}(\emph{t})
\emph{e}^{-\lambda_{BS}\pi(r_{\emph{k}_{\emph{i}_{UL}}}^{RRU-UE}(\emph{t}))^2} \hfill \\
\rm \exp(-{\frac{(\emph{e}^\omega-1)\cdot(I_{mean-RR{U_\emph{k}}}(\emph{t})+N_0+p_{T-RR{U_\emph{k}}}(\emph{t}, \emph{e}_\emph{k})\cdot h_{self}(\emph{t}))\cdot r_{\emph{k}_{\emph{i}_{UL}}}^{RRU-UE}(\emph{t})^\alpha}
{p_{T-RR{U_\emph{k}}}(\emph{t}, \emph{e}_\emph{k})}})dr_{\emph{k}_{\emph{i}_{UL}}}^{RRU-UE}(\emph{t})d\omega \hfill
\end{gathered}.\quad\tag{24d}\
\]
\hrulefill
\vspace*{4pt}
\end{figure*}
In the end,  $\rm EE_{\emph{k}}^{UL-mean}(\emph{t})$ is expressed as (25).And the $\Phi_2$ is illustrated in (24d).
\[\rm EE_{\emph{k}}^{UL-mean}(\emph{t}) = {\frac{\Phi_2}
{p_{T-UE}(\emph{t}, \emph{e}_\emph{k})+p_{UE-static}}}, \tag{25}\]%bad box

Based on the results of $(23)$ and $(25)$, the weighted energy efficiency of the BS $\rm BS_\emph{k}$, {\em i.e.,}
$\rm U_{BS_\emph{k}}^{mean}(p_{T-RR{U_\emph{k}}}(\emph{t}, \emph{e}_\emph{k}), m(\emph{t}, \emph{e}))$ in the mean field $\rm m(\emph{t}, \emph{e})$ of network energy efficiency is expressed as (26). $\Phi_1$ and $\Phi_2$ is illustrated in (22d) and (24d).

\[\begin{aligned}
&\rm U_{BS_\emph{k}}^{mean}(p_{T-RR{U_\emph{k}}}(\emph{t}, \emph{e}_\emph{k}), m(\emph{t}, \emph{e})) \hfill \\
&\rm= EE_{\emph{k}}^{DL-mean}(\emph{t})+\beta\cdot EE_{\emph{k}}^{UL-mean}(\emph{t})\hfill \\
&\rm = {\frac{\Phi_1}
{p_{T-RR{U_\emph{k}}}(\emph{t}, \emph{e}_\emph{k})+p_{static}}} \\
\quad\rm &+\beta\cdot{\frac{\Phi_2}
{p_{T-UE}(\emph{t}, \emph{e}_\emph{k})+p_{UE-static}}}
\end{aligned},
\tag{26}\]
\subsection{Mean Field Game Equations}
\label{sec3-3}
When the differential game is replaced by the mean field game to optimize the energy efficiency of cellular networks,  the differential game of RRU transmission power strategy $\rm \varphi_G$ is replaced by the mean field game of RRU transmission power strategy $\rm \varphi_{G-mean}$.

\textbf{Definition 5 (Mean field game of RRU transmission power strategy $\rm \varphi_{G-mean}$ ):}
Every BS is assumed as a player for the cellular network. The utility function of the player $k$ is expressed as:
\[\rm UF_\emph{k}^{mean} = E[{\int_0^\emph{T}}U_{BS_\emph{k}}^{mean}(p_{T-RR{U_\emph{k}}}(\emph{t}, \emph{e}_\emph{k}), m(\emph{t}, \emph{e}))d\emph{t}], \tag{27}\]
with the object function:
\[\rm \max\limits_{p_{T-RR{U_\emph{k}}}(0\rightarrow \emph{T})}UF_\emph{k}^{mean}\]
$$
\begin{aligned}
\mathrm{s.t.}\quad&\rm 0\leq p_{T-RR{U_\emph{k}}}(\emph{t}, \emph{e}_\emph{k})\leq p_{T-RR{U_{max}}},  \\
\quad&\rm d\emph{e}_\emph{k}(\emph{t})=-p_{T-RR{U_\emph{k}}}(\emph{t}, \emph{e}_\emph{k})d\emph{t}, 0\leq \emph{t}\leq \emph{T}.\\
\end{aligned}
$$

Based on the properties of mean field game \cite{Guea40}\cite{Schulte41}, the solution of Nash equilibrium $\rm \varphi_{G-mean}$
can be derived by (28).
\begin{figure*}[!t]
\[\left\{ \begin{gathered}
\rm \frac{\partial V_\emph{k}(\emph{t}, \emph{e}_\emph{k})}{\partial \emph{t}}+\max\limits_{p_{T-RR{U_\emph{k}}}(\emph{t}, \emph{e}_\emph{k})}(U_{BS_\emph{k}}^{mean}(p_{T-RR{U_\emph{k}}}(\emph{t}, \emph{e}_\emph{k}), m(\emph{t}, \emph{e}))
-p_{T-RR{U_\emph{k}}}(\emph{t}, \emph{e}_\emph{k})\cdot\frac{\partial V_\emph{k}(\emph{t}, \emph{e}_\emph{k})}{\partial \emph{e}})=0 \hfill \\
\rm \frac{\partial m(\emph{t}, \emph{e}_\emph{k})}{\partial \emph{t}}-\frac{\partial}{\partial \emph{t}}(m(\emph{t}, \emph{e}_\emph{k})p_{T-RR{U_\emph{k}}}(\emph{t}, \emph{e}_\emph{k}))=0 \hfill \\
\end{gathered}.\right.\tag{28}\]
\end{figure*}
To derive the Nash equilibrium of mean field game $\rm \varphi_{G-mean}$, every BS needs to solve (28). The RRU transmission power strategy of every BS is the same in the mean field game model \cite{Lasry20}\cite{Guea40}. Moreover,  every point is independent and similar in the Poisson point process \cite{Andrews39}. Therefore, the solution of Nash equilibrium with mean field game of RRU transmission power strategy $\rm \varphi_{G-mean}$ is the same for every BS in the cellular network. Based on the mean field game theory, (28) can be simplified as (29a) and (29b).
\begin{figure*}
\setcounter{equation}{28}
\begin{subnumcases}{}
 \rm \frac{\partial V(\emph{t}, \emph{e})}{\partial \emph{t}}+\max\limits_{p(\emph{t}, \emph{e})}(U_{BS}^{mean}(p(\emph{t}, \emph{e}), m(\emph{t}, \emph{e}))
-p(\emph{t}, \emph{e})\cdot\frac{\partial V(\emph{t}, \emph{e})}{\partial \emph{e}})=0, \hfill                    \\
\rm  \frac{\partial m(\emph{t}, \emph{e})}{\partial \emph{t}}-\frac{\partial}{\partial \emph{t}}(m(\emph{t}, \emph{e})p(\emph{t}, \emph{e}))=0,  \hfill
\end{subnumcases}
\end{figure*}
$\rm p(\emph{t}, \emph{e})$ is the RRU transmission power at a BS,  $\rm U_{BS}^{mean}(p(\emph{t}, \emph{e}), m(\emph{t}, \emph{e}))$  is the weighted energy efficiency at a BS, $\rm V(\emph{t}, \emph{e})$ is the value function of a BS.
(29a) and (29b) are the HJB and Fokker-Planck-Kolmogorov (FPK) equations respectively. Based on the solutions of (29a) and (29b), the Nash equilibrium point $\rm p_{mean}^*$ and the mean filed $\rm m^*$ at the Nash equilibrium point can be obtained for the mean field game of RRU transmission power strategy $\rm \varphi_{G-mean}$.

\section{Algorithm Design of Mean Field Game}
Although there is no analytical solution for (29a) and (29b) \cite{Lasry20}\cite{Guea40}, we can obtain the numerical solutions of (29a) and (29b) by the finite difference method and the Lax-Friedrichs method \cite{Ahmed42}. The time period $\rm T$ is discretized into $(X+1)$ time points,  {\em i.e., }the time point set
$\rm \emph{t}=\{0, 1\ast\delta \emph{t}, 2\ast\delta \emph{t}, 3\ast\delta \emph{t}, ..., \emph{X}\ast\delta \emph{t}\}$, where $\delta t=\frac{T}{X}$. The maximum volume of battery at the RRU is discretized into $(Y+1)$ energy points,  {\em i.e.,} the energy point set
$e=\{0, 1\ast\delta e, 2\ast\delta e, 3\ast\delta e, ..., Y\ast\delta e\}$, where $\rm \delta e=\frac{E_{max}}{\emph{Y}}$. Based on the time point set and the energy point set, $\rm V(\emph{t}, \emph{e})$ and $\rm m(\emph{t}, \emph{e})$ are expressed by $\rm V(\emph{x}, \emph{y})$ and $\rm m(\emph{x}, \emph{y})$, where $0\leq x\leq X$ and $0\leq y\leq Y$.
\subsection{Numerical Solution of FPK and HJB}
\label{sec2-1}
Based on the Lax-Friedrichs method \cite{Ahmed42}, (29b) is extended as (30).
\begin{figure*}[!t]
\[\rm m(\emph{x}+1, \emph{y}) = \frac{1}{2}[m(\emph{x}, \emph{y}-1)+m(\emph{x}, \emph{x}+1)]+\frac{\delta \emph{t}}{2(\delta \emph{e})}[p(\emph{x}, \emph{x}+1)m(\emph{x}, \emph{x}+1)-p(\emph{x}, \emph{x}-1)m(\emph{x}, \emph{x}-1)].\tag{30}\]
\end{figure*}
Utilizing the discretization method,  $\rm \frac{\partial V(\emph{t}, \emph{e})}{\partial \emph{t}}$ and $\rm \frac{\partial V(\emph{t}, \emph{e})}{\partial \emph{e}}$ are expressed as (31) and (32),
\[\rm \frac{\partial V(\emph{t}, \emph{e})}{\partial \emph{t}} = \frac{V(\emph{x}, \emph{y})-V(\emph{x}-1, \emph{y})}{\delta \emph{t}},\tag{31}\]
%Utilizing the discretization method,  $\frac{\partial V(t, e)}{\partial e}$  is expressed as
\[\rm \frac{\partial V(\emph{t}, \emph{e})}{\partial \emph{e}} = \frac{V(\emph{x}, \emph{y})-V(\emph{x}, \emph{y}-1)}{\delta \emph{e}}.\tag{32}\]

Substitute (31) and (32) into (29a),  we have the (33).
\begin{figure*}
\[\rm \frac{V(\emph{x}, \emph{y})-V(\emph{x}-1, \emph{y})}{\delta \emph{t}}+\max\limits_{p(\emph{x}, \emph{y})}(U_{BS}^{mean}(p(\emph{x}, \emph{y}), m(\emph{x}, \emph{y}))-p(\emph{x}, \emph{y})\cdot  \frac{V(\emph{x}, \emph{y})-V(\emph{x}, \emph{y}-1)}{\delta \emph{e}}) = 0.\tag{33}\]
\hrulefill
\vspace*{4pt}
\end{figure*}
\subsection{Algorithm Design for Numerical Solution of Mean Field Game}
\label{sec2-2}
Based on (30) and (33),  the equilibrium of mean field game (EMFG) Algorithm is developed to obtain the Nash equilibrium $\rm p_{mean}^*$ and the mean filed $\rm m^*$ at the Nash equilibrium.
\rule{0.49\textwidth}{0.2mm}

\textbf{Equilibrium of Mean Field Game Algorithm (EMFG Algorithm)}\\
\rule{0.49\textwidth}{0.2mm}
Initialization: Initialize $\rm V(\emph{X}, :)$, $\rm m(:, :)$\\
Repeat:

{\setlength\parindent{1em} {For $\emph{x}=X:-1:0$ do}}

    {\setlength\parindent{2em}For $\emph{y}=Y:-1:0$ do}

      {\setlength\parindent{3em}Solving:}

      {\setlength\parindent{4em}$\rm \max\limits_{p(\emph{x}, \emph{y})}(U_{BS}^{mean}(p(\emph{x}, \emph{y}), m(\emph{x}, \emph{y}))-$}

      {\setlength\parindent{5em} $\rm p(\emph{x}, \emph{y})\cdot \frac{V(\emph{x}, \emph{y})-V(\emph{x}, \emph{y}-1)}{\delta \emph{e}})$, get $p(\emph{x}, \emph{y})$}

    {\setlength\parindent{2em} End for

    Update $\rm V(\emph{x}-1, :)$ using:

    $\rm \frac{V(\emph{x}, \emph{y})-V(\emph{x}-1, \emph{y})}{\delta \emph{t}}+\max\limits_{p(\emph{x}, \emph{y})}(U_{BS}^{mean}(p(\emph{x}, \emph{y}), m(\emph{x}, \emph{y}))-p(\emph{x}, \emph{y})\cdot$

    $\frac{V(\emph{x}, \emph{y})-V(\emph{x}, \emph{y}-1)}{\delta \emph{e}}) = 0$

  \setlength\parindent{1em}End for

  Update $\rm m(:, :)$ using:

  $\rm m(\emph{x}+1, \emph{y})=\frac{1}{2}[m(\emph{x}, \emph{y}-1)+m(\emph{x}, \emph{y}+1)]+$

  $\frac{\delta \emph{t}}{2(\delta \emph{e})}[p(\emph{x}, \emph{y}+1)m(\emph{x}, \emph{y}+1)-p(\emph{x}, \emph{y}-1)m(\emph{x}, \emph{y}-1)]$\\
Until Convergence\\
Obtain $\rm p_{mean}^*=p(\emph{x}, \emph{y})$, $\rm m^*=m(\emph{x}, \emph{y})$\\
Output: $\rm p_{mean}^*$, $\rm m^*$\\
\rule{0.49\textwidth}{0.2mm}
\section{Numerical Simulations of Mean Field Game}
Based on the EMFG Algorithm, the value of $\rm V(\emph{t}, \emph{e})$ at the end of the period $T$ needs to be configured in advance. To save of battery energy at RRUs, the value of $\rm V(\emph{t}, \emph{e})$ at the end of the period $T$ is configured as $\rm V(\emph{X}, \emph{y}\ast\delta \emph{e}) = 0.05\ast\exp(\emph{y}\ast\delta \emph{e})$ in this paper. Moreover, the residual battery volume of RRUs is assumed to be governed by a uniform distribution at the initial stage of simulations, {\em i.e., }$\rm m(:, :)=\frac{1}{\emph{X}+1}$. The detail parameters of numerical simulations are list in the Table II.
\begin{table}[!hbt]
\centering
\caption{\ Parameters for Numerical Simulations}
\begin{tabular}{|cccc|}
\hline
    \textbf{Parameter} & \textbf{Default value}
    & \textbf{Parameter} & \textbf{Default value}\\
    $\rm {p_{static}}$&${6}$W&${\alpha}$&${3}$\\
    ${\rm \lambda_{BS}}$&${0.0005/m^{2}}$&${\rm N_0}$&${10^{-8}}$W\\
    ${\rm p_{T-UE}}$&${0.1}$W&${\rm p_{UE-static}}$&${0.5}$W\\
    ${\rm h_{self}}$&${0.0004}$&${\rm p_{T-RR{U_{max}}}}$&
    ${1}$W\\
    ${\rm E_{max}}$&${2}$J&${T}$&${1}$S\\
    \hline
    \end{tabular}
\label{tab1}
\end{table}

\begin{figure}[!h]%3
\centering
\includegraphics[width=0.5\textwidth]{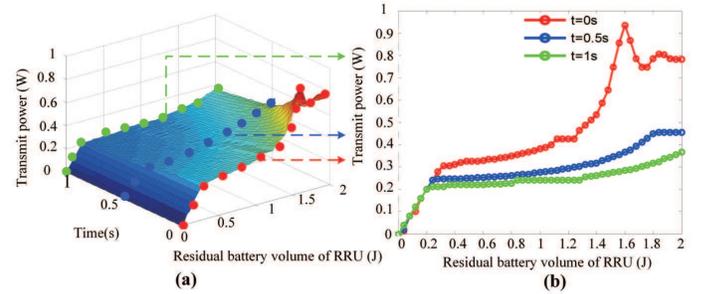}
\caption{RRU transmission power with respect to the time and residual battery volume of RRU}
\label{Fig 3}
\end{figure}

In Fig. 3, we provide the result of the simulation for the proposed algorithms to gain insight.
When the weighted factor for uplinks is configured as $\beta = 1$, the RRU transmission power with respect to the time and residual battery volume of RRU is illustrated in Fig. 3. From Fig. 3(a),  the RRU transmission power decreases with increasing time and decreasing residual battery volume at RRU. To clearly explain the results in Fig. 3 (a),  three time slots, {\em i.e., }$t = 0$, $t = 0.5$ and $t = 1$ are selected and plotted in Fig. 3(b). When the time slot is fixed in Fig. 3(b),  the RRU transmission power increases with increasing the residual battery volume of RRU. When the residual battery volume of RRU is fixed in Fig. 3(b),  the RRU transmission power decreases with increasing the time slot.

\begin{figure}[!h]%4
\centering
\includegraphics[width=3.5in, draft=false]{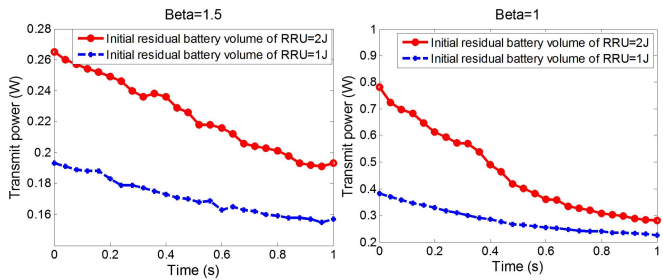}
\caption{RRU transmission power with respect to different initial battery volumes of RRU and weighted factors for uplinks}
\end{figure}

Fig. 4 shows the RRU transmission power with respect to different initial battery volumes of RRU and weighted factors for uplinks. When the weighted factor is fixed,  the RRU transmission power with initial residual battery volume 2 J is larger than or equal to the RRU transmission power with initial residual battery volume 1 J. When the initial residual battery volume is fixed,  the RRU transmission power decreases with increasing the weighted factor for uplinks.

\begin{figure}[!h]%5
\centering
\includegraphics[width=0.5\textwidth]{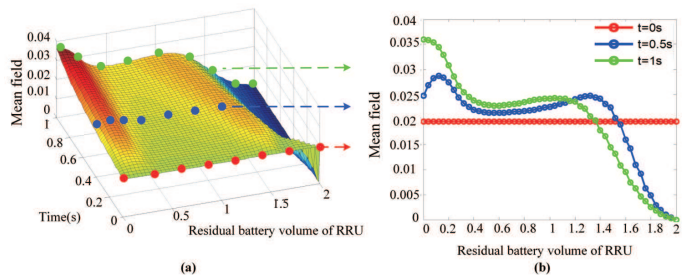}
\caption{Mean field with respect to the residual battery volume and the time}
\end{figure}

To analyze the distribution of residual battery volumes of RRUs in the cellular network,  the mean field with respect to the residual battery volume and the time is depicted in Fig. 5. Without loss of generality,  the weighted factor of uplinks is configured as $\beta = 1$. Based on the results in Fig. 5(a),  the mean field with the low residual battery volume of RRU increases with the increase of time. To clearly explain the results in Fig. 5(a),  three time slots, {\em i.e.,} $t = 0$, $t = 0.5$ and $t = 1$ are selected and plotted in Fig. 5(b). When the time slot $t = 0$ is ignored and the time slot is fixed in Fig. 5(b),  the mean field decreases with increasing the residual battery volume of RRU. When the residual battery volume is less than or equal to 1.2 J in Fig. 5(b),  the mean field with the time slot $t = 1$ is larger than the mean field with the time slots $t = 0$ and $t = 0.5$. When the residual battery volume is larger than or equal to 1.6 J in Fig. 5(b),  the mean field with the time slot $t = 0$ is larger than the mean field with the time slots $t = 0.5$ and $t = 1$. These results indicate that the number of RRUs with high residual battery volume, {\em i.e., }the residual battery volume is larger than or equal to 1.6 J,  is larger than the number of RRUs with low residual battery volume,  i.e.,  the residual battery volume is less than 1.6 J in the initial time. In the end of a period $T$, the number of RRUs with low residual battery volume, {\em i.e., }the residual battery volume is less than or equal to 1.2 J,  is larger than the number of RRUs with high residual battery volume, {\em i.e., }the residual battery volume is larger than 1.2 J.

\begin{figure}[!h]%6
\centering
\includegraphics[width=3.5in, draft=false]{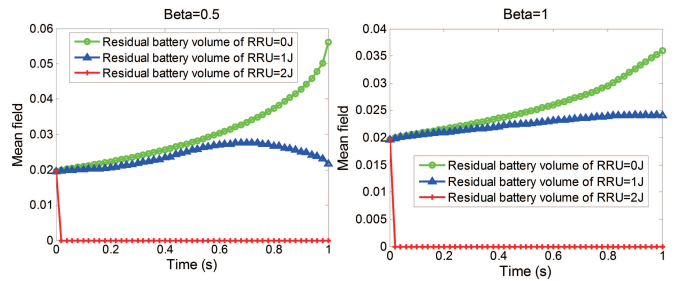}
\caption{Mean field with respect to the time and the weighted factor for uplinks}
\end{figure}

When the initial residual battery volume of RRUs are configured as 0,  1 and 2 J,  the mean field with respect to the time and the weighted factor for uplinks is shown in Fig. 6. When the time is fixed, the mean field with the initial residual battery volume 0 J is larger than or equal to the mean field with the initial residual battery volume 1 and 2 J. When the residual battery volume is fixed as 0 J, the mean field increases with the increase of time. When the residual battery volume is fixed as 2 J,  the mean field quickly falls into zero with increasing the time. When the residual battery volume is fixed as 1 J and the weighted factor $\beta$ is fixed as 0.5,  the mean field first increases with increasing the time and then the mean field decreases with increasing the time after the time is larger than 0.7 second (s). The reason of this change is that the number of RRUs with the residual battery volume 1 J is added by the RRUs with the initial residual battery volume 2 J in the start time, {\em i.e.,} the time is less than 0.7s. When the time is larger than or equal to 0.7s, the number of RRUs with the residual battery volume 1 J is reduced by the energy consuming for RRU transmission power. Moreover, few RRUs with the initial residual battery volume 2 J is added into the number of RRUs with the residual battery volume 1 J. When the residual battery volume is fixed as 1 J and the weighted factor $\beta$ is fixed as 0.5 and 1,  the mean field increases with increasing the time.

To evaluate the performance of EMFG Algorithm,  the network energy efficiency is defined as
\[\rm \Psi(\emph{t}) = \sum\limits_{\emph{e}}m(\emph{t}, \emph{e})\cdot U_{BS}^{mean}(p(\emph{t}, \emph{e}), m(\emph{t}, \emph{e})).\tag{34}\]

Without loss of generality,  the initial RRU transmission power is configured as 1 Watt (W) in cellular networks. The network energy efficiencies adopted the fixed RRU transmission power strategy and the RRU transmission power strategy of mean field game, {\em i.e.,} the EMFG algorithm, are compared in Fig. 7. Considering different weighted factors,  the network energy efficiency under the RRU transmission power strategy of mean field game is always larger than that under the fixed RRU transmission power strategy. Moreover, the network energy efficiency under the RRU transmission power strategy of mean field game increases with the increase of the time in 5G cellular networks. To the contrary,  the network energy efficiency adopted the fixed RRU transmission power strategy decreases with the increase of the time in 5G cellular networks. Hence, the RRU transmission power strategy of mean field game, {\em i.e.,} the proposed EMFG algorithm can improve the network energy efficiency of 5G cellular networks.

\begin{figure}[!h]%8
\centering
\includegraphics[width=3.5in, draft=false]{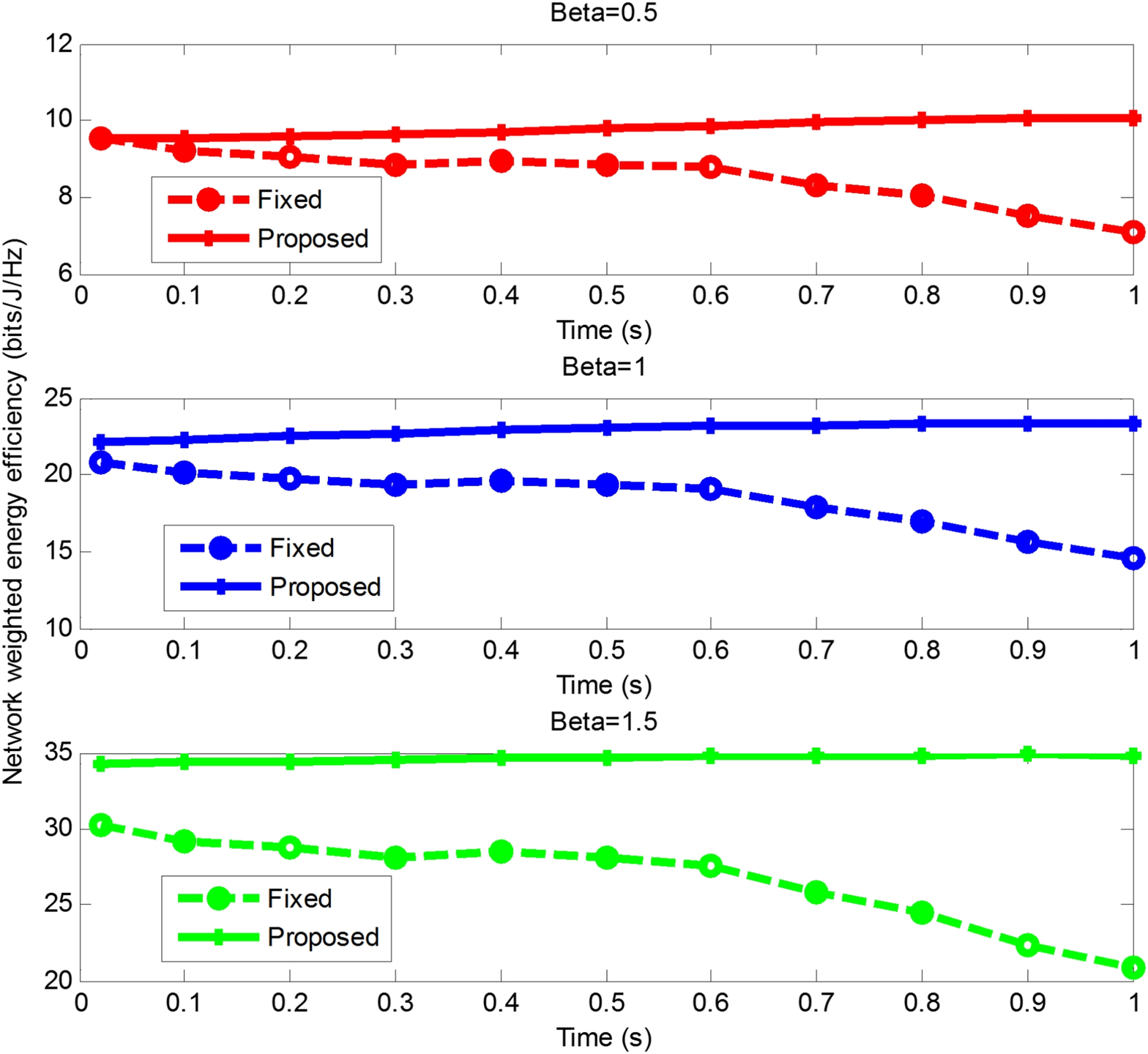}
\caption{Weighted network energy efficiency with respect to the time compared with the fixed RRU transmission power strategy and the RRU transmission power of mean field game strategy}
\end{figure}

The coverage probability is another key metric of the RRU transmission power strategy for 5G cellular networks. The coverage probability of RRU is denoted as $\rm E[Pr(SINR_{\emph{k}_\emph{i}}^{DL}(\emph{t})>\delta)]$, where $\delta$ is the threshold of receive signal power at UEs. The average network coverage probability is expressed as
\[\rm \Theta = \frac{1}{N_{RRU}}\sum\limits_{\emph{k}=1}^{N_{RRU}}E[Pr(SINR_{\emph{k}_\emph{i}}^{DL}(\emph{t})>\delta)], \tag{35}\]
where $\rm N_{RRU}$ is the number of RRUs in cellular networks. The average network coverage probability with respect to the time considering different thresholds of receive signal power at UEs are compared for cellular networks with the fixed RRU transmission power strategy and the RRU transmission power strategy of mean field game in Fig. 8. For different thresholds of receive signal power at UEs,  {\em e.g.,} $\delta$ is -10,  0 and 10 dB,  the average network coverage probability with the RRU transmission power strategy of mean field game always keeps a stable level with increasing the time and the average network coverage probability with the fixed RRU transmission power strategy decreases with increasing the time. In the initial time, the average network coverage probability with the fixed RRU transmission power strategy is larger than the average network coverage probability with the RRU transmission power strategy of mean field game in 5G cellular networks. After certain time,  {\em e.g.,} 0.1, 0 and 0.3 seconds corresponding to different thresholds of receive signal power at UEs, {\em i.e.,} -10, 0 and 10 dB,  the average network coverage probability with the fixed RRU transmission power strategy is less than that with the RRU transmission power strategy of mean field game in 5G cellular networks. Therefore, the RRU transmission power strategy of mean field game, {\em i.e.,} the proposed EMFG algorithm, is helpful for keeping the stable of average network coverage probability in a long time scale.

\begin{figure}[!h]%9
\centering
\includegraphics[width=3.5in, draft=false]{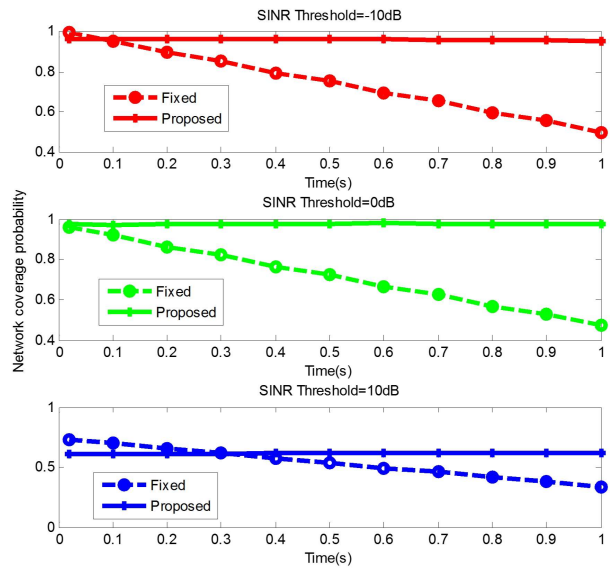}
\caption{Average network coverage probability with respect to the time compared with the fixed RRU transmission power strategy and the RRU transmission power of mean field game strategy}
\end{figure}

\section{Conclusions}
In this paper we investigate the network energy efficiency of 5G full duplex cellular networks by the mean field game theory. Based on the solution of mean field game equations, a new RRU transmission power strategy, {\em i.e.,} the EMFG algorithm is developed to optimize the network energy efficiency of 5G full duplex cellular networks. Simulation results show that the network energy efficiency of the proposed EMFG algorithm is larger than the network energy efficiency of the fixed RRU transmission power algorithm in 5G full duplex cellular networks. Moreover, the cellular network adopted the proposed EMFG algorithm is helpful for keeping the average network coverage probability at a stable level. For the future work, we plan to investigate the network energy efficiency with quality of service constraints for 5G full duplex cellular networks based on the mean field game theory.

%\section*{Acknowledgment}

\ifCLASSOPTIONcaptionsoff
  \newpage
\fi

\end{document}